\documentclass[a4paper,12pt]{article}
\usepackage[T2A]{fontenc} 
\usepackage[russian, english]{babel} 
\usepackage[dvips]{graphicx} 
\makeindex 
\usepackage{epsfig} 
\usepackage[dvips]{texdraw} \input{txdtools}
\usepackage{picinpar} 
\usepackage{amsmath}
\usepackage{amssymb}
\usepackage{indentfirst} 

\begin{document}
\begin{center}

{\bf \large $K^+_{l3\gamma}$ decays revisited:\\
\vspace{2mm} branching ratios and $T$-odd momenta correlations}

\vspace{5mm}

I.B.\,Khriplovich\footnote{khriplovich@inp.nsk.su} and
A.S.\,Rudenko\footnote{a.s.rudenko@inp.nsk.su}

Budker Institute of Nuclear Physics,\\ 630090 Novosibirsk, Russia

\end{center}

\begin{abstract}
We calculate the branching ratios of the $K^+ \to \pi^0 l^+ \nu_l\gamma \hspace{2mm}
(l = e, \mu)$ decays, and the $T$-odd triple momenta correlations
$\xi=\vec{q}\cdot[\vec{p}_l \times \vec{p}_\pi]/M^3_K$, due to the electromagnetic
final state interaction, in these processes. The contributions on the order of
$\omega^{-1}$ and $\omega^0$ to the corresponding amplitudes are treated exactly.
For the branching ratios, the corrections on the order of $\omega$ are estimated and
demonstrated to be small. We compare the results with those of other authors. In
some cases our results differ considerably from the previous ones.

\end{abstract}

\vspace{5mm}

{\bf 1.} The $K^+ \to \pi^0 l^+ \nu_l\gamma \hspace{2mm} (l = e, \mu)$ decays were
earlier studied theoretically in \linebreak Refs.~\cite{gas,brag,kub}. Therein the
branching ratios of these decays were calculated. Besides, in Ref.~\cite{brag} the
$T$-odd triple momenta correlations $\xi=\vec{q}\cdot[\vec{p}_l \times
\vec{p}_\pi]/M^3_K$ were considered, as induced by the electromagnetic final state
interactions; here and below $M_K$ is the kaon mass, $\vec{q}$, $\vec{p}_l$,
$\vec{p}_\pi$ are the momenta of $\gamma$, $l^+$, $\pi^0$, respectively. In
principle, these triple correlations can be used to probe new $CP$-odd effects
beyond the Standard Model, which could also contribute to them.

Here we calculate anew these effects. Our results confirm essentially some previous
results and disagree considerably with other ones.

In the theoretical analysis of radiative effects in the discussed processes, the
treatment of the accompanying radiation, which gives the effects on the order of
$\omega^{-1}$ and $\omega^0$ (the last ones originate from the radiation due to the
lepton magnetic moment), is straightforward (here and below $\omega$ is the photon
energy). As to the structure radiation contribution on the order of $\omega^0$, it
is also under control, due in fact to the gauge invariance \cite{low}. The
contributions on the order of $\omega$ (and higher) depend directly on the photon
field strength $F_{\mu\nu}$ (and its derivatives), and cannot be fixed in a
model-independent way. We assume that the corrections on the order of $\omega$ and
higher are relatively small. And indeed, more quantitative arguments presented below
demonstrate that such contributions into the discussed branching ratios do not
exceed few percent.


\vspace{3mm}

\begin{figure}[h]
\center
\begin{tabular}{c c c c c}
\includegraphics[scale=1.2]{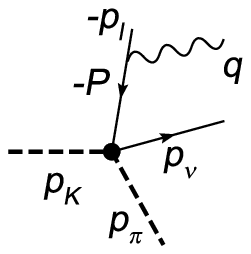} &  &
\includegraphics[scale=1.2]{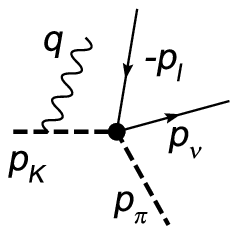} &  &
\includegraphics[scale=1.2]{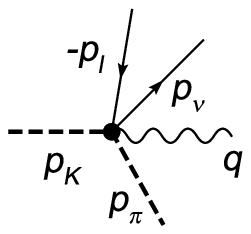} \\
a & & b & & c
\end{tabular}
\caption {The tree diagrams} \label{fig:1}
\end{figure}

{\bf 2.} At the tree level, the $K^+ \to \pi^0 l^+ \nu_l \gamma$ decay is described
by the Feynman graphs in Fig.~\ref{fig:1}.

The matrix elements for diagrams 1a and 1b look as follows:
\begin{multline}\label{M1a}
M_{1a}=\frac{G}{\sqrt{2}}\sin{\theta_c}e\bar{u}_\nu\gamma_\alpha(1+\gamma_5)\frac{\hat{p_l}+\hat{q}-m_l}{2p_lq}\hat{e}^*v_l[f_+(t)\cdot(p_K+p_\pi)_\alpha+f_-(t)\cdot(p_K-p_\pi)_\alpha]\\
=\frac{G}{\sqrt{2}}\sin{\theta_c}e\bar{u}_\nu\gamma_\alpha(1+\gamma_5)\left(\frac{2p_le^*+\hat{q}\hat{e}^*}{2p_lq}\right)v_l[f_+(t)\cdot(p_K+p_\pi)_\alpha+f_-(t)\cdot(p_K-p_\pi)_\alpha],
\end{multline}
\begin{equation}\label{M1b}
M_{1b}=-\frac{G}{\sqrt{2}}\sin{\theta_c}e\bar{u}_\nu\gamma_\alpha(1+\gamma_5)v_l[f_+(t^\prime)
\cdot(p_K-q+p_\pi)_\alpha+f_-(t^\prime)\cdot(p_K-q-p_\pi)_\alpha]\frac{p_Ke^*}{p_Kq};
\end{equation}

\noindent here $G$ is the Fermi coupling constant, $\theta_c$ is the Cabibbo angle,
$e$ is the elementary charge ($e>0$), $t=(p_K-p_\pi)^2$, $t^\prime=(p_K-q-p_\pi)^2$;
the lower indices attached to the matrix elements match the corresponding Feynman
diagrams.

Usually the dependence of the form factors $f_+$ and $f_-$ on the momentum transfer $t$ is described by
formula
\begin{equation} \label{f-}
f_\pm(t)=f_\pm(0)\left(1+\lambda_\pm\frac{t}{m^2_\pi}\right).
\end{equation}
The experimental data are adequately described by Eq.~(\ref{f-}) with
$\lambda_+\approx 0.03$ for $l=\mu$ and $l=e$; $\lambda_-=0$ for $l=\mu$
\cite{PDG08}; $\lambda_-$ for $l=e$ is unknown, but one may assume that it is also
close to zero.

In the $K^+_{l3\gamma}$ decays the ratio $\lambda_\pm t/m^2_\pi$ is small,
$\lambda_\pm t/m^2_\pi \lesssim 0.1$, so one can put $f_\pm(t)=f_\pm(0)$. Since the
ratio $\xi(0)=f_-(0)/f_+(0)\sim 0.1$ is also small \cite{PDG04}, one can neglect
$f_-(0)$ with the same accuracy.

Thus, our expressions (\ref{M1a}), (\ref{M1b}) simplify to

\begin{equation}
M_{1a}=\frac{G}{\sqrt{2}}\sin{\theta_c}ef_+(0)\cdot(p_K+p_\pi)_\alpha\bar{u}_\nu\gamma_\alpha(1+\gamma_5)\left(\frac{p_le^*}{p_lq}+\frac{\hat{q}\hat{e}^*}{2p_lq}\right)v_l,
\end{equation}

\begin{equation}
M_{1b}=-\frac{G}{\sqrt{2}}\sin{\theta_c}ef_+(0)\cdot(p_K-q+p_\pi)_\alpha\frac{p_Ke^*}{p_Kq}\bar{u}_\nu\gamma_\alpha(1+\gamma_5)v_l.
\end{equation}

However, the sum of diagrams 1a and 1b is not gauge invariant: it does not vanish
under the substitution $e^*\to q$. To restore the gauge invariance, one should add
the third diagram where a photon is directly emitted from the vertex (see
Fig.~\ref{fig:1}c). This contact amplitude has no single-particle intermediate
states, and therefore is on the order of $\omega^0$ and higher. The contribution
$\sim\omega^0$, as derived with the Low technique \cite{low}, is

\begin{equation}
M_{1c}=-\frac{G}{\sqrt{2}}\sin{\theta_c}ef_+(0)\cdot
e^*_\alpha\bar{u}_\nu\gamma_\alpha(1+\gamma_5)v_l.
\end{equation}

Thus, the model-independent gauge invariant tree amplitude of $K^+_{l3\gamma}$
decay, including only terms on the order of $\omega^{-1}$ and $\omega^0$ (but all of
them!), is
\begin{multline} \label{m1}
M_1=M_{1a}+M_{1b}+M_{1c}=\frac{G}{\sqrt{2}}\sin{\theta_c}ef_+(0)\left\{(p_K+p_\pi)_\alpha\bar{u}_\nu\gamma_\alpha(1+\gamma_5)v_l\left(\frac{p_le^*}{p_lq}-\frac{p_Ke^*}{p_Kq}\right) \right. \\
\left.
+(p_K+p_\pi)_\alpha\bar{u}_\nu\gamma_\alpha(1+\gamma_5)\frac{\hat{q}\hat{e}^*}{2p_lq}v_l+\left(\frac{p_Ke^*}{p_Kq}q_\alpha-e^*_\alpha\right)\bar{u}_\nu\gamma_\alpha(1+\gamma_5)v_l\right\}.
\end{multline}
This expression agrees with the corresponding formulas in Ref.~\cite{gas} and
formula (6) in Ref.~\cite{brag} (if our $f_+(0)$ is set to its $SU(3)$ value
$f_+(0)=1/\sqrt{2}$\,).

It is convenient to present the amplitude (\ref{m1}) as a sum of gauge invariant
contributions. They are the "infrared" term $M_{IR}$ corresponding to the sum of the
amplitudes of accompanying radiation by the kaon and lepton (independent of the
lepton magnetic moment), the magnetic term $M_{mag}$ which is the amplitude of
spin-dependent accompanying radiation of the lepton magnetic moment, and the Low
term $M_{Low}$:

\begin{equation}
M_{IR}=\frac{G}{\sqrt{2}}\sin{\theta_c}ef_+(0)(p_K+p_\pi)_\alpha\bar{u}_\nu\gamma_\alpha(1+\gamma_5)v_l\left(\frac{p_le^*}{p_lq}-\frac{p_Ke^*}{p_Kq}\right),
\end{equation}

\begin{equation}
M_{mag}=\frac{G}{\sqrt{2}}\sin{\theta_c}ef_+(0)(p_K+p_\pi)_\alpha\bar{u}_\nu\gamma_\alpha(1+\gamma_5)\frac{\hat{q}\hat{e}^*}{2p_lq}v_l,
\end{equation}

\begin{equation}
M_{Low}=\frac{G}{\sqrt{2}}\sin{\theta_c}ef_+(0)\left(\frac{p_Ke^*}{p_Kq}q_\alpha-e^*_\alpha\right)\bar{u}_\nu\gamma_\alpha(1+\gamma_5)v_l.
\end{equation}

The results of calculation for $K^+_{l3\gamma}$ branching ratios are presented in
Table~\ref{table:1}; here the following cuts in the kaon rest frame are used:
 $\omega \geqslant 30$ MeV
and $\theta_{l\gamma}\geqslant 20^\circ$.
\begin{table*}[h]
\begin{center}
\renewcommand{\arraystretch}{1.3}
\begin{tabular}{|c|c|c|} \hline
  & $l=\mu$ & $l=e$ \\
\hline
Bijnens et al. \cite{gas} & $1.9\times 10^{-5}$ & $2.8\times 10^{-4}$\\
\hline Braguta et al. \cite{brag} & $2.15\times
10^{-5}$ & $3.18\times 10^{-4}$ \\
\hline
present work & $1.81\times 10^{-5}$&$2.72\times 10^{-4}$\\
\hline experimental values& $(2.4\pm0.5\pm0.6)\times
10^{-5}$ \cite{shim} &$(3.06\pm0.09\pm0.14)\times 10^{-4}$ \cite{akim}\\
& $(1.58\pm0.46\pm0.08)\times 10^{-5}$ \cite{adler} &\\\hline
\end{tabular}
\caption{Branching ratio of $K^+ \to \pi^0 l^+ \nu_l \gamma$ decays} \label{table:1}
\end{center}
\end{table*}

The accuracy of our results can be estimated as follows. The leading corrections to
them are due to the structure radiation from the hadronic vertex. They are
proportional to the photon field strength, i.e.\ are on the order of $\omega$. There
are good reasons to believe that these corrections are less than the Low structure
amplitudes which are on the order of $\omega^0$. The Low contributions (including of
course their interference with the accompanying radiation) to the discussed $K^+ \to
\pi^0 \mu^+ \nu_\mu \gamma$ and $K^+ \to \pi^0 e^+ \nu_e \gamma$ branching ratios,
according to our calculations, constitute $- 0.24 \times 10^{-5}$ and
$-0.12\times10^{-4}$, respectively. Thus, we estimate the accuracy of our results as
$\pm 0.2 \times 10^{-5}$ and $\pm 0.1 \times 10^{-4}$, correspondingly. Let us note
also that corrections to the quoted results derived in Ref.~\cite{gas,kub} in the
chiral perturbation theory are of similar magnitude.

As mentioned, additional corrections on the level of 10 -- 15\% to the branching
ratios originate from our neglect of the form factor $f_-(t)$ and of the
$t$-dependence of $f_+(t)$ (the last correction is certainly positive). As to the
relative accuracy of our numerical integration over phase space of final particles,
it is about 1\%.

To compare properly our results with those of Refs.~\cite{gas,brag} one should keep
in mind that now the experimental values of some quantities are known with better
accuracy. Indeed, we use $\sin{\theta_c}f_+(0)=0.217/\sqrt{2}$ in our calculation,
and, as far as we can see, in Refs.~\cite{gas,brag} the corresponding value is
$0.22/\sqrt{2}$. Substitution of one of these values for another alters the results
by about $3\%$.

Thus, our results for the branching ratios agree reasonably well with those of
Ref.~\cite{gas}. There is however some disagreement between our results and those of Ref.~\cite{brag}.

\vspace{1cm}

\begin{figure}[h]
\center
\begin{tabular}{c c c c c}
\includegraphics[scale=1.2]{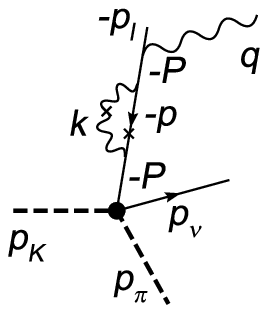} & &
\includegraphics[scale=1.2]{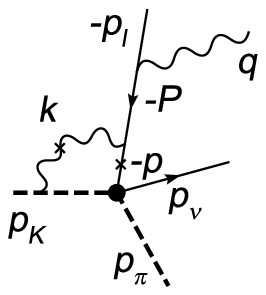} & &
\includegraphics[scale=1.2]{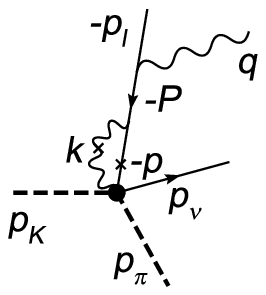} \\
a & & c & & e \\
& & & & \\
\includegraphics[scale=1.2]{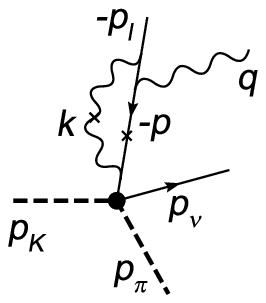} & &
\includegraphics[scale=1.2]{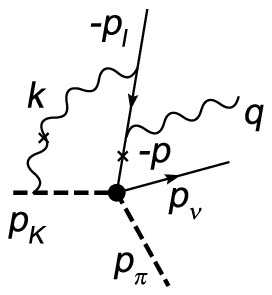} & &
\includegraphics[scale=1.2]{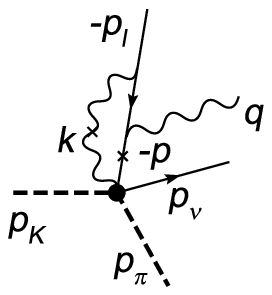} \\
b & & d & & f
\end{tabular}
\caption {The one-loop diagrams} \label{fig:2}
\end{figure}

{\bf 3.} The $T$-odd  triple momenta correlation $\xi=\vec{q}\cdot[\vec{p}_l \times
\vec{p}_\pi]/M^3_K$ in the $K^+ \to \pi^0 l^+ \nu_l \gamma$ decays arise from the
interference term $2\mathrm{Re}(M^*_1 A_2)$ in the decay rate; here $M_1$ is the
tree amplitude and $A_2$ is the anti-Hermitian part of the one-loop diagrams
presented in Fig.~\ref{fig:2}.

One can easily demonstrate that $A_2$ is generated only by attaching the
intermediate photon to diagram 1a. The on-mass-shell intermediate particles in
Fig.~\ref{fig:2} are marked by crosses.

The anti-Hermitian part of the sum of one-loop diagrams is written as

\begin{equation} \label{A2}
A_2=\frac{i}{8\pi^2}\sum_n M_{fn}M^*_{in}.
\end{equation}

\begin{figure}[h]
\center
\begin{tabular}{c c}
\includegraphics[scale=1.2]{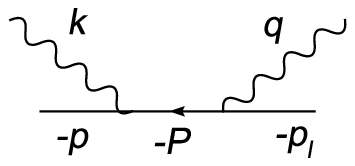} &
\includegraphics[scale=1.2]{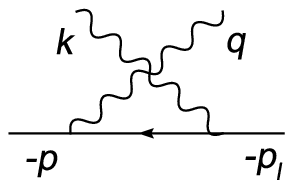} \\
a & b
\end{tabular}
\caption {The Compton scattering diagrams} \label{fig:3}
\end{figure}

The Compton amplitude entering this expression is (see Fig.~\ref{fig:3})
\begin{equation}
M_{fn}=M_{3a}+M_{3b}=e^2\bar{v}_p \hat{e}_k \frac{\hat{P}-m_l}{2p_{l}q} \hat{e}^*
v_l + e^2\bar{v}_p \hat{e}^* \frac{\hat{p}_l-\hat{k}-m_l}{-2p_{l}k} \hat{e}_k v_l.
\end{equation}

As to $M^*_{in}$, it is the same tree amplitude (\ref{m1}), up to the change of some
notations:
\begin{multline} \label{m1kp}
M^*_{in}=M_{ni}=M_{1}\rvert_{\substack{q\to k,\\ p_{l}\to p,\\ e\to e_k}}=\frac{G}{\sqrt{2}}\sin{\theta_c}ef_+(0)\left\{(p_K+p_\pi)_\alpha\bar{u}_\nu\gamma_\alpha(1+\gamma_5)v_p\left(\frac{pe^*_k}{p_{l}q}-\frac{p_Ke^*_k}{p_Kk}\right) \right. \\
\left.
+(p_K+p_\pi)_\alpha\bar{u}_\nu\gamma_\alpha(1+\gamma_5)\frac{\hat{k}\hat{e}^*_k}{2p_{l}q}v_p+
\left(\frac{p_Ke^*_k}{p_Kk}k_\alpha-e^*_{k\alpha}\right)\bar{u}_\nu\gamma_\alpha(1+\gamma_5)v_p\right\}.
\end{multline}
The sum over $n$ in formula (\ref{A2}) includes the summation over the polarizations
of intermediate particles, and the integral over the phase space with
\begin{equation}
d\rho=\frac{d^3k}{2\omega_k}\frac{d^3p}{2E_p}\delta^{(4)}(p+k-P).
\end{equation}

The details of the calculation of the $T$-odd correlation
$\xi=\vec{q}\cdot[\vec{p}_l \times \vec{p}_\pi]/M^3_K$ are given in Appendix. Here
we wish to note only that our result for $\xi$-odd term $|M|^2_{odd}$ in the
interference $2\mathrm{Re}(M^*_1 A_2)$ agrees with formula (12) in Ref.~\cite{brag}
up to the factor $1/\sqrt{2}$ which is obviously omitted therein (it follows
directly from formulas (6) and (10) of Ref.~\cite{brag}).

In fact, what is really measured experimentally, is not the $T$-odd triple momentum
correlation $\xi$ by itself, but the asymmetry
\begin{equation}
\begin{split}
A_\xi&=\frac{N_+-N_-}{N_++N_-}=\frac{\int
\left(|M_1|^2+|M|^2_{odd}\right) d\Phi_{\xi>0}-\int
\left(|M_1|^2+|M|^2_{odd}\right) d\Phi_{\xi<0}}{\int
\left(|M_1|^2+|M|^2_{odd}\right) d\Phi_{\xi>0}+\int
\left(|M_1|^2+|M|^2_{odd}\right) d\Phi_{\xi<0}} \\
&=\frac{\int |M|^2_{odd}d\Phi_{\xi>0}}{\int |M_1|^2d\Phi_{\xi>0}}
\end{split}
\end{equation}
induced by this correlation; here $N_+$ and $N_-$ are the numbers of events with
$\xi>0$ and $\xi<0$, and integration is performed over the phase space of the final
particles.

The results for the asymmetry $A_\xi$ are presented in Table~\ref{table:2}. The
relative accuracy of our numerical integration is about 1\%, as it was the case with
the branching ratios.

\begin{table*}[h]
\begin{center}
\renewcommand{\arraystretch}{1.3}
\begin{tabular}{|c|c|c|} \hline
& $l=\mu$ & $l=e$ \\\hline Braguta et al. \cite{brag} & $1.14\times 10^{-4}$ & $-0.59\times 10^{-4}$ \\
& $\omega
> 30$ MeV, $\theta_{l\gamma}> 20^\circ$ & $\omega
> 30$ MeV, $\theta_{l\gamma}> 20^\circ$ \\
\hline present work & $2.38\times 10^{-4}$ & $-0.93\times 10^{-4}$ \\ & $\omega
\geqslant 30$ MeV, $\theta_{l\gamma}\geqslant 20^\circ$ & $\omega \geqslant 30$
MeV, $\theta_{l\gamma}\geqslant 20^\circ$ \\
\hline experimental values & $-0.03\pm0.13$~\cite{tchik} & $-0.015\pm0.021$~\cite{akim} \\
 & $5<\omega<30$ MeV & $\omega>10$ MeV, $0.6<\cos\theta_{l\gamma}<0.9$ \\
\hline present work & $0.50\times 10^{-4}$ & $-0.30\times
10^{-4}$ \\
 & $5<\omega<30$ MeV &$\omega>10$ MeV, $0.6<\cos\theta_{l\gamma}<0.9$ \\
\hline
\end{tabular}
\caption{$A_\xi$ in $K^+ \to \pi^0 l^+ \nu_l \gamma$ decay}\label{table:2}
\end{center}
\end{table*}

Here however, as distinct from the problem of branching ratios, the contribution of
the Low term is large, quite comparable numerically to the contributions of the
accompanying radiation which are on the order of $\omega^{-1}$ and $\omega^0$. So,
here it is difficult to estimate reliably the relative magnitude of the structure
radiation contribution proportional to $\omega$, i.e.\ to estimate reliably the true
accuracy of thus derived results for the asymmetry $A_\xi$. We note here that
corrections to the value of the discussed correlation, derived in Ref.~\cite{mul}
within the chiral perturbation theory, are very small.

Our results for $A_\xi$ exceed those of Ref.~\cite{brag}, obtained in the same
approximation, by a factor of 1.5 -- 2. If one included the factor $1/\sqrt{2}$,
probably lost in the calculations of Ref.~\cite{brag} (see the remark above), it
would make the disagreement even worse.

The asymmetry $A_\xi$ was measured experimentally~\cite{akim,tchik}, but cuts
imposed therein differ from those used in Ref.~\cite{brag}. Therefore, we have
calculated also the asymmetry for the corresponding kinematical regions (see the
last line in Table~\ref{table:2}).

As distinct from the situation with the branching ratios, all the theoretical
results for the triple correlations are unfortunately far away from the real
experimental sensitivity.

\subsection*{Acknowledgements}

We are grateful to A.E.\,Bondar and L.B.\,Okun for their interest to the work and
useful discussions, to L.V.\,Kardapoltsev for his advices concerning numerical
calculations, to V.V.\,Braguta for the discussion of the cuts used in
Ref.~\cite{brag}, and to B.\,Kubis for attracting our attention to
Refs.~\cite{kub,mul}.

The work was supported in part by the Russian Foundation for Basic Research through
Grant No.\ 08-02-00960-a and by the Federal Program "Personnel of Innovational
Russia" through Grant No.\ 14.740.11.0082.

\appendix
\subsection*{Appendix A}

\setcounter{equation}{0}
\renewcommand{\theequation}{A.\arabic{equation}}

In this section we give the list of the integrals that contribute to the $A_2$:

\begin{equation}
\int d\rho=a_0=\frac{\pi}{2}\left(1-\frac{m^2_l}{P^2}\right);
\end{equation}

\begin{equation}
\int k_\mu d\rho=a_P P_\mu, \mspace{5mu} \mathrm{where} \mspace{10mu}
a_P=\frac{\pi}{4}\left(1-\frac{m^2_l}{P^2}\right)^2;
\end{equation}

\begin{equation}
\int \frac{1}{p_lk}
d\rho=b_0=\frac{\pi}{2p_lq}\mathrm{ln}\left(\frac{P^2}{m^2_l}\right);
\end{equation}

\begin{equation}
\int \frac{k_\mu}{p_lk} d\rho=B_{1\mu}=b_{l}p_{l\mu}+b_{P}P_\mu\,,
\end{equation}
here $b_{l}$ and $b_{P}$ are the solutions of the system of equations
\begin{align*}
b_{l}m^2_l+b_{P}(p_lP)&=a_0,\\
b_{l}(p_lP)+b_{P}P^2&=b_0(p_lq);
\end{align*}

\begin{equation}
\int \frac{k_\mu k_\nu}{p_lk} d\rho=B_{2\mu \nu}=b_{2}g_{\mu
\nu}+b_{ll}p_{l\mu}p_{l\nu}+b_{PP}P_\mu P_\nu+b_{lP}(p_{l\mu}P_\nu+P_\mu p_{l\nu}),
\end{equation}
here $b_2$, $b_{ll}$, $b_{PP}$, and $b_{lP}$ are the solutions of the system of
equations
\begin{align*}
4b_2+b_{ll}m^2_l+b_{PP}P^2+2b_{lP}(p_lP) &= 0,\\
b_2+b_{ll}m^2_l+b_{lP}(p_lP) &= 0,\\
b_{PP}(p_lP)+b_{lP}m^2_l &= a_P,\\
b_2+b_{PP}P^2+b_{lP}(p_lP) &= b_{P}(p_lq);
\end{align*}

\begin{equation}
\int \frac{1}{p_Kk}
d\rho=c_0=\frac{\pi}{2}\frac{1}{\sqrt{(p_KP)^2-m^2_KP^2}}\;\mathrm{ln}\!\!\left(\frac{p_KP+\sqrt{(p_KP)^2-m^2_KP^2}}{p_KP-\sqrt{(p_KP)^2-m^2_KP^2}}\right);
\end{equation}

\begin{equation}
\int \frac{k_\mu}{p_Kk} d\rho=C_{1\mu}=c_{K}p_{K\mu}+c_{P}P_\mu,
\end{equation}
here $c_{K}$ and $c_{P}$ are the solutions of the system of equations
\begin{align*}
c_{K}m^2_K+c_{P}(p_KP)&=a_0,\\
c_{K}(p_KP)+c_{P}P^2&=c_0(p_lq);
\end{align*}

\begin{equation}
\int \frac{1}{(p_lk)(p_Kk)}
d\rho=d_0=\frac{\pi}{2p_lq}\frac{1}{\sqrt{(p_Kp_l)^2-m^2_Km^2_l}}\;\mathrm{ln}\!\!\left(\frac{p_Kp_l+\sqrt{(p_Kp_l)^2-m^2_Km^2_l}}{p_Kp_l-\sqrt{(p_Kp_l)^2-m^2_Km^2_l}}\right);
\end{equation}

\begin{equation}
\int \frac{k_\mu}{(p_lk)(p_Kk)}
d\rho=D_{1\mu}=d_{K}p_{K\mu}+d_{l}p_{l\mu}+d_{P}P_\mu,
\end{equation}
here $d_{K}$, $d_{l}$, and $d_{P}$ are the solutions of the system of equations
\begin{align*}
d_{K}m^2_K+d_{l}(p_Kp_l)+d_{P}(p_KP)&=b_0,\\
d_{K}(p_Kp_l)+d_{l}m^2_l+d_{P}(p_lP)&=c_0,\\
d_{K}(p_KP)+d_{l}(p_lP)+d_{P}P^2&=d_0(p_lq);
\end{align*}

\begin{multline}
\int \frac{k_\mu k_\nu}{(p_lk)(p_Kk)} d\rho=D_{2\mu \nu}=d_{2}g_{\mu
\nu}+d_{KK}p_{K\mu}p_{K\nu}+d_{ll}p_{l\mu}p_{l\nu}+d_{PP}P_\mu
P_\nu\\+d_{Kl}(p_{K\mu}p_{l\nu}+p_{l\mu} p_{K\nu})+d_{KP}(p_{K\mu}P_\nu+P_\mu
p_{K\nu})+d_{lP}(p_{l\mu}P_\nu+P_\mu p_{l\nu}),
\end{multline}
here $d_2$, $d_{KK}$, $d_{ll}$, $d_{PP}$, $d_{Kl}$, $d_{KP}$, and $d_{lP}$ are the
solutions of the system of equations
\begin{align*}
4d_2+d_{KK}m^2_K+d_{ll}m^2_l+d_{PP}P^2+2d_{Kl}(p_Kp_l)+2d_{KP}(p_KP)+2d_{lP}(p_lP)&=0,\\
d_2+d_{ll}m^2_l+d_{Kl}(p_Kp_l)+d_{lP}(p_lP)&=0,\\
d_{KK}(p_Kp_l)+d_{Kl}m^2_l+d_{KP}(p_lP)&=c_{K},\\
d_{PP}(p_lP)+d_{KP}(p_Kp_l)+d_{lP}m^2_l&=c_{P},\\
d_2+d_{KK}m^2_K+d_{Kl}(p_Kp_l)+d_{KP}(p_KP)&=0,\\
d_{PP}(p_KP)+d_{KP}m^2_K+d_{lP}(p_Kp_l)&=b_{P},\\
d_2+d_{PP}P^2+d_{KP}(p_KP)+d_{lP}(p_lP)&=d_{P}(p_lq).
\end{align*}

All these expressions agree with the analogous ones in Ref.~\cite{brag}.

\subsection*{Appendix B}

\setcounter{equation}{0}
\renewcommand{\theequation}{B.\arabic{equation}}

The $A_2$ can be expressed as follows:

\begin{multline}
A_2=\frac{i}{8\pi^2}\sum_n M_{fn}M^*_{in}=\frac{i}{8\pi^2}\sum_n
(M_{3a}+M_{3b})(M_{IR}+M_{mag}+M_{Low})\rvert_{\substack{q\to k,\\
p_{l}\to p,\\ e\to
e_k}}\\=A_{3a-IR}+A_{3a-mag}+A_{3a-Low}+A_{3b-IR}+A_{3b-mag}+A_{3b-Low},
\end{multline}
where
\begin{multline}
A_{3a-IR}=\frac{i}{8\pi^2}\sum_n
M_{3a}M_{IR}\rvert_{\substack{q\to k,\\
p_{l}\to p,\\ e\to
e_k}}=\frac{ie^2}{8\pi^2}\frac{G}{\sqrt{2}}\sin{\theta_c}ef_+(0)(p_K+p_\pi)_\alpha\bar{u}_\nu\gamma_\alpha(1+\gamma_5)\\
\times
\left[c_0(\hat{P}-m_l)\hat{p}_K-\hat{C}_1\hat{p}_K+\frac{m_l}{p_lq}(a_0(\hat{P}-m_l)-a_P\hat{P})\right]\frac{\hat{P}-m_l}{2p_lq}
\hat{e}^* v_l;
\end{multline}

\begin{multline}
A_{3a-mag}=\frac{i}{8\pi^2}\sum_n
M_{3a}M_{mag}\rvert_{\substack{q\to k,\\
p_{l}\to p,\\ e\to
e_k}}=\frac{ie^2}{8\pi^2}\frac{G}{\sqrt{2}}\sin{\theta_c}ef_+(0)(p_K+p_\pi)_\alpha\bar{u}_\nu\gamma_\alpha(1+\gamma_5)\\
\times \left[2a_P(P^2+2m_l\hat{P})\right]\frac{\hat{P}-m_l}{(2p_lq)^2} \hat{e}^*
v_l;
\end{multline}

\begin{multline}
A_{3a-Low}=\frac{i}{8\pi^2}\sum_n
M_{3a}M_{Low}\rvert_{\substack{q\to k,\\
p_{l}\to p,\\ e\to
e_k}}=\frac{ie^2}{8\pi^2}\frac{G}{\sqrt{2}}\sin{\theta_c}ef_+(0)\bar{u}_\nu\gamma_\alpha(1+\gamma_5)\\
\times
\left[a_0(\hat{P}-m_l)\gamma_\alpha-a_P\hat{P}\gamma_\alpha-C_{1\alpha}(\hat{P}-m_l)\hat{p}_K\right]\frac{\hat{P}-m_l}{2p_lq}
\hat{e}^* v_l;
\end{multline}

\begin{multline}
A_{3b-IR}=\frac{i}{8\pi^2}\sum_n
M_{3b}M_{IR}\rvert_{\substack{q\to k,\\
p_{l}\to p,\\ e\to
e_k}}=\frac{ie^2}{8\pi^2}\frac{G}{\sqrt{2}}\sin{\theta_c}ef_+(0)(p_K+p_\pi)_\alpha\bar{u}_\nu\gamma_\alpha(1+\gamma_5)\\
\times
\frac{1}{2}\left[(\hat{P}-m_l)\hat{e}^*\left(\frac{1}{p_lq}((\hat{p}_l-m_l)(b_0\hat{P}-\hat{B}_1)-
\hat{B}_1\hat{P})-d_0(\hat{p}_l-m_l)\hat{p}_K+
\hat{D}_1\hat{p}_K\right)\right.\\
\left.-\frac{1}{p_lq}(\hat{B}_1\hat{e}^*(\hat{p}_l-m_l)\hat{P}-
B_{2\mu\nu}\gamma_\mu\hat{e}^*(\hat{p}_l-m_l)\gamma_\nu-B_{2\mu\nu}\gamma_\mu\hat{e}^*\gamma_\nu\hat{P})
\right.\\
\left.+\hat{D}_1\hat{e}^*(\hat{p}_l-m_l)\hat{p}_K-D_{2\mu\nu}\gamma_\mu\hat{e}^*\gamma_\nu\hat{p}_K\right]v_l;
\end{multline}

\begin{multline}
A_{3b-mag}=\frac{i}{8\pi^2}\sum_n
M_{3b}M_{mag}\rvert_{\substack{q\to k,\\
p_{l}\to p,\\ e\to
e_k}}=\frac{ie^2}{8\pi^2}\frac{G}{\sqrt{2}}\sin{\theta_c}ef_+(0)(p_K+p_\pi)_\alpha\bar{u}_\nu\gamma_\alpha(1+\gamma_5)\\
\times
\frac{1}{2p_lq}\left[B_{2\mu\nu}\gamma_\mu\hat{p}_l\hat{e}^*\gamma_\nu+4m_lB_{2\mu\nu}\gamma_\mu
e^*_\nu-\hat{B}_1\hat{p}_l\hat{e}^*\hat{P}-4m_l\hat{B}_1(p_le^*)-m^2_l\hat{B}_1\hat{e}^*\right]v_l;
\end{multline}

\begin{multline}
A_{3b-Low}=\frac{i}{8\pi^2}\sum_n
M_{3b}M_{Low}\rvert_{\substack{q\to k,\\
p_{l}\to p,\\ e\to
e_k}}=\frac{ie^2}{8\pi^2}\frac{G}{\sqrt{2}}\sin{\theta_c}ef_+(0)\bar{u}_\nu\gamma_\alpha(1+\gamma_5)\\
\times
\frac{1}{2}\left[(\hat{P}-m_l)\hat{e}^*((\hat{p}_l-m_l)(D_{1\alpha}\hat{p}_K-b_0\gamma_\alpha)-D_{2\alpha\mu}\gamma_\mu\hat{p}_K+
\hat{B}_1\gamma_\alpha)\right.\\
\left.+\hat{B}_1\hat{e}^*(\hat{p}_l-m_l)\gamma_\alpha-B_{2\mu\nu}\gamma_\mu\hat{e}^*\gamma_\nu\gamma_\alpha\right]v_l.
\end{multline}

\subsection*{Appendix C}

\setcounter{equation}{0}
\renewcommand{\theequation}{C.\arabic{equation}}

The interference term $2\mathrm{Re}(M^*_1 A_2)$ can be expressed
as follows:
\begin{multline}
2\mathrm{Re}(M^*_1 A_2)=2\mathrm{Re}((M^*_{IR}+M^*_{mag}+M^*_{Low})\\
\times
(A_{3a-IR}+A_{3a-mag}+A_{3a-Low}+A_{3b-IR}+A_{3b-mag}+A_{3b-Low})).
\end{multline}

We keep here only terms odd in $\xi$. Let the $|M|^2_{IR-3a-IR}$ be the $\xi$-odd
terms in $2\mathrm{Re}(M^*_{IR} A_{3a-IR})$, the $|M|^2_{IR-3a-mag}$ be the
$\xi$-odd terms in $2\mathrm{Re}(M^*_{IR} A_{3a-mag})$, etc.

\vspace{5mm}

Therefore, the $\xi$-odd term $|M|^2_{odd}$ in the interference term
$2\mathrm{Re}(M^*_1 A_2)$ is

\begin{equation}
\begin{split}
|M|^2_{odd}&=|M|^2_{IR-3a-IR}+|M|^2_{IR-3a-mag}+|M|^2_{IR-3a-Low}+|M|^2_{IR-3b-IR}\\
&+|M|^2_{IR-3b-mag}+|M|^2_{IR-3b-Low}+|M|^2_{mag-3a-IR}+|M|^2_{mag-3a-mag}\\
&+|M|^2_{mag-3a-Low}+|M|^2_{mag-3b-IR}+|M|^2_{mag-3b-mag}+|M|^2_{mag-3b-Low}\\
&+|M|^2_{Low-3a-IR}+|M|^2_{Low-3a-mag}+|M|^2_{Low-3a-Low}+|M|^2_{Low-3b-IR}\\
&+|M|^2_{Low-3b-mag}+|M|^2_{Low-3b-Low},
\end{split}
\end{equation}

where
\begin{equation}
\begin{split}
|M|^2_{IR-3a-IR}&=\left(\frac{G}{\sqrt{2}}\sin{\theta_c}ef_+(0)\right)^2 \frac{e^2
\xi m^4_K}{\pi^2 (p_Kq) (p_lq)^2}\\ &\times
(3m^2_K-m^2_{\pi}-2(p_Kp_l-p_Kp_\pi+p_Kq+p_lp_\pi+p_\pi q))\\ &\times
((2a_0-a_P)m^2_l+c_{K}m^2_K(p_lq)-c_{P}m^2_l(p_Kq)+2(c_{P}-c_0)(p_Kp_l+p_Kq)(p_lq));
\end{split}
\end{equation}

\begin{equation}
\begin{split}
|M|^2_{IR-3a-mag}&=\left(\frac{G}{\sqrt{2}}\sin{\theta_c}ef_+(0)\right)^2 \frac{e^2
\xi m^4_K}{\pi^2 (p_Kq) (p_lq)^2}\\ &\times
a_P(m_l^2-2p_lq)(3m^2_K-m^2_{\pi}-2(p_Kp_l-p_Kp_\pi+p_Kq+p_lp_\pi+p_\pi q));
\end{split}
\end{equation}

\begin{equation}
\begin{split}
|M|^2_{IR-3a-Low}&=\left(\frac{G}{\sqrt{2}}\sin{\theta_c}ef_+(0)\right)^2 \frac{2e^2
\xi m^4_K}{\pi^2 (p_Kq)(p_lq)^2}\\ &\times
((c_{K}+2c_{P})((p_lq)^2m_K^2+(p_Kq)^2m_l^2+(p_Kp_l)(p_Kq)m_l^2)\\
&+2c_{K}(p_Kp_l+p_Kq)(p_lq)m_K^2-c_{P}(p_Kp_l+p_Kq)(p_lq)(2p_lq+m_l^2)\\
&-2c_{K}((p_Kp_l)^2+(p_Kq)^2)(p_lq)-4c_{K}(p_Kp_l)(p_Kq)(p_lq)\\
&+(a_0+a_P)(p_lq-2p_Kq)m_l^2+2(a_P-a_0)(p_lq)^2);
\end{split}
\end{equation}

\begin{equation}
\begin{split}
|M|^2_{IR-3b-IR}&=\left(\frac{G}{\sqrt{2}}\sin{\theta_c}ef_+(0)\right)^2 \frac{e^2
\xi m^4_K}{\pi^2 (p_Kq) (p_lq)}\\ &\times
(-3m^2_K+m^2_{\pi}+2(p_Kp_l-p_Kp_\pi+p_Kq+p_lp_\pi+p_\pi q))\\ &\times
(2b_2+(2b_0-5b_{l}-5b_{P}+b_{ll}+4b_{lP}+3b_{PP})m_l^2\\
&+2(b_0-b_{l}-3b_{P}+b_{lP}+2b_{PP})(p_lq)+(d_{K}-2d_{KP})(p_lq)m_K^2\\
&+(-d_{l}-3d_{P}+2d_{lP}+2d_{PP})m_l^2(p_Kq)+2d_{KP}(p_Kp_l)(p_Kq)\\
&+2(-d_0+d_{l}+2d_{P}-d_{lP}-d_{PP})(p_Kp_l)(p_lq));
\end{split}
\end{equation}

\begin{equation}
\begin{split}
|M|^2_{IR-3b-mag}&=\left(\frac{G}{\sqrt{2}}\sin{\theta_c}ef_+(0)\right)^2 \frac{e^2
\xi m^4_K}{\pi^2 (p_Kq) (p_lq)}\\ &\times
(-3m^2_K+m^2_{\pi}+2(p_Kp_l-p_Kp_\pi+p_Kq+p_lp_\pi+p_\pi q))\\
&\times((b_{l}+b_{P}-2b_{lP}-2b_{PP})m_l^2+2(b_{P}-b_{PP})(p_lq));
\end{split}
\end{equation}

\begin{equation}
\begin{split}
|M|^2_{IR-3b-Low}&=\left(\frac{G}{\sqrt{2}}\sin{\theta_c}ef_+(0)\right)^2 \frac{-e^2
\xi m^4_K}{\pi^2 (p_Kq)(p_lq)}\\ &\times
(-2d_{KK}(p_lq)m_K^4+2(d_{Kl}-2d_{PP})(p_lq)^2m_K^2\\
&+(-d_{KK}+4d_{Kl}+4d_{KP})(p_Kq)m_l^2m_K^2+2(d_{ll}-d_{PP})(p_lq)m_l^2m_K^2\\
&+2(2d_{K}+d_{KK}-4d_{Kl}-2d_{KP})(p_Kp_l)(p_lq)m_K^2+2d_{KK}(p_Kq)(p_lq)m_K^2\\
&+2(-d_{Kl}-3d_{KP})(p_Kq)^2m_l^2+4(-2b_{P}+b_{lP}+b_{PP})(p_lq)^2\\
&+4(-d_{l}-d_{P}+d_{ll}+2d_{lP}+d_{PP})(p_Kp_l)(p_lq)^2\\
&+(-d_{ll}-4d_{lP}-3d_{PP})(p_Kq)m_l^4+2(-6b_{P}+4b_{lP}+4b_{PP}-3d_2)(p_Kq)m_l^2\\
&+2(d_{K}-2d_{l}+2d_{P}-d_{Kl}-2d_{KP}+2d_{ll}+4d_{lP}+2d_{PP})(p_Kp_l)(p_Kq)m_l^2\\
&+2(-3b_{P}+b_{ll}+2b_{lP}+b_{PP})(p_lq)m_l^2\\
&+4(-d_{K}+2d_{l}+d_{Kl}+d_{KP}-2d_{ll}-2d_{lP})(p_Kp_l)^2(p_lq)\\
&+4b_2(p_lq)+2(-d_{l}-d_{P}+d_{ll}+2d_{lP}+d_{PP})(p_Kp_l)(p_lq)m_l^2\\
&+4(4b_{P}-2b_{lP}-2b_{PP}+d_2)(p_Kp_l)(p_lq)\\
&+2(-d_{ll}-3d_{lP}-3d_{PP})(p_Kq)(p_lq)m_l^2\\
&+4(-d_{K}+d_{Kl}+2d_{KP}+2d_{PP})(p_Kp_l)(p_Kq)(p_lq)\\
&+4(b_0-b_{l})(p_Kq)m_l^2+2(b_0-3b_{l})(p_lq)m_l^2+4(b_0-b_{l})(p_lq-2p_Kp_l)(p_lq));
\end{split}
\end{equation}

\begin{equation}
\begin{split}
|M|^2_{mag-3a-IR}&=\left(\frac{G}{\sqrt{2}}\sin{\theta_c}ef_+(0)\right)^2 \frac{e^2
\xi m^4_K}{\pi^2 (p_lq)^2}(c_{P}(m_l^2-2p_lq)+2c_0(p_lq))\\
&\times(3m^2_K-m^2_{\pi}-2(p_Kp_l-p_Kp_\pi+p_Kq+p_lp_\pi+p_\pi q));
\end{split}
\end{equation}

\begin{equation}
|M|^2_{mag-3a-mag}=0;
\end{equation}

\begin{equation}
\begin{split}
|M|^2_{mag-3a-Low}&=\left(\frac{G}{\sqrt{2}}\sin{\theta_c}ef_+(0)\right)^2
\frac{-2e^2 \xi m^4_K}{\pi^2 (p_lq)^2}\\ &\times
(2c_{K}(p_lq)m_K^2-2c_{P}(p_lq)^2+(c_{K}+2c_{P})(p_Kp_l+p_Kq)m_l^2\\
&-c_{P}(p_lq)m_l^2-2c_{K}(p_Kp_l+p_Kq)(p_lq)+4(a_P-a_0)(p_lq)-2(a_0+a_P)m_l^2);
\end{split}
\end{equation}

\begin{equation}
\begin{split}
|M|^2_{mag-3b-IR}&=\left(\frac{G}{\sqrt{2}}\sin{\theta_c}ef_+(0)\right)^2 \frac{e^2
\xi m^4_K}{\pi^2 (p_lq)}\\ &\times
(3m^2_K-m^2_{\pi}-2(p_Kp_l-p_Kp_\pi+p_Kq+p_lp_\pi+p_\pi
q))\\
&\times(d_{KK}m_K^2+(d_{ll}+2d_{lP}+d_{PP}-d_{l}-d_{P})m_l^2+2(d_{Kl}+d_{KP}-d_{K})(p_Kp_l));
\end{split}
\end{equation}

\begin{equation}
|M|^2_{mag-3b-mag}=0;
\end{equation}

\begin{equation}
\begin{split}
|M|^2_{mag-3b-Low}&=\left(\frac{G}{\sqrt{2}}\sin{\theta_c}ef_+(0)\right)^2
\frac{-e^2 \xi m^4_K}{\pi^2 (p_lq)}\\ &\times
((d_{ll}+2d_{lP}+d_{PP})m_l^4+4(2b_{l}+2b_{P}-2b_{ll}-4b_{lP}-2b_{PP}+d_2)m_l^2\\
&+d_{KK}m_K^2m_l^2+2(-d_{K}-2d_{l}-2d_{P}+d_{Kl}+d_{KP})(p_Kp_l)m_l^2\\
&+2(d_{Kl}+2d_{KP}+2d_{ll}+4d_{lP}+2d_{PP})(p_Kq)m_l^2\\
&+2(d_{lP}+d_{PP})(p_lq)m_l^2-8b_2+4b_0(m_l^2-2p_lq)\\
&-2(d_{KK}+2d_{Kl}+2d_{KP})(p_lq)m_K^2+8(b_{l}+2b_{P}-b_{lP}-b_{PP})(p_lq)\\
&+4(d_{K}+2d_{l}+2d_{P}-d_{Kl}-d_{KP}-2d_{ll}-4d_{lP}-2d_{PP})(p_Kp_l)(p_lq));
\end{split}
\end{equation}

\begin{equation}
\begin{split}
|M|^2_{Low-3a-IR}&=\left(\frac{G}{\sqrt{2}}\sin{\theta_c}ef_+(0)\right)^2 \frac{2e^2
\xi m^4_K}{\pi^2 (p_Kq)(p_lq)^2}\\ &\times
(a_0(p_lq)m_l^2+2(a_P-2a_0)(p_Kp_l)m_l^2+(c_{K}-2c_0+2c_{P})(p_lq)^2m_K^2\\
&-2c_{K}(p_Kp_l)(p_lq)m_K^2+2(c_0-c_{P})(p_Kp_l)(p_lq)(2p_Kp_l+2p_Kq-p_lq)\\
&+c_{P}m_l^2(p_Kp_l)(p_lq));
\end{split}
\end{equation}

\begin{equation}
\begin{split}
|M|^2_{Low-3a-mag}&=\left(\frac{G}{\sqrt{2}}\sin{\theta_c}ef_+(0)\right)^2 \frac{e^2
\xi m^4_K}{\pi^2 (p_Kq)(p_lq)^2}\\ &\times
(-2a_P)(2(p_Kp_l)m_l^2+(p_lq)m_l^2+2(p_lq)^2-4(p_Kp_l)(p_lq));
\end{split}
\end{equation}

\begin{equation}
\begin{split}
|M|^2_{Low-3a-Low}&=\left(\frac{G}{\sqrt{2}}\sin{\theta_c}ef_+(0)\right)^2
\frac{2e^2 \xi m^4_K}{\pi^2(p_Kq)(p_lq)}(2(a_P-a_0)(p_lq)\\
&-2c_{P}(p_Kp_l)(p_lq)+c_{K}((p_lq)m_K^2-2(p_Kp_l)^2-2(p_Kp_l)(p_Kq)));
\end{split}
\end{equation}

\begin{equation}
\begin{split}
|M|^2_{Low-3b-IR}&=\left(\frac{G}{\sqrt{2}}\sin{\theta_c}ef_+(0)\right)^2 \frac{e^2
\xi m^4_K}{\pi^2 (p_Kq)(p_lq)}\\ &\times
(-2d_{KK}(p_lq)m_K^4+2(-d_{K}-2d_{P}+2d_{KP}+2d_{lP}+2d_{PP})(p_lq)^2m_K^2\\
&+2(-2d_{l}-2d_{P}+d_{Kl}+d_{KP}+d_{ll}+2d_{lP}+d_{PP})(p_lq)m_l^2m_K^2\\
&+4d_2(p_lq)m_K^2+8(d_{K}-d_{Kl}-d_{KP})(p_Kp_l)(p_lq)m_K^2+2d_{KK}(p_Kq)(p_lq)m_K^2\\
&+(2b_{l}+2b_{P}-3b_{ll}-6b_{lP}-3b_{PP})m_l^4+4(b_{l}-b_{ll}-b_{lP})(p_Kq)m_l^2\\
&+4(-b_0+b_{l}+3b_{P}-b_{lP}-2b_{PP})(p_lq)^2\\
&+4(d_0-d_{l}-2d_{P}+d_{lP}+d_{PP})(p_Kp_l)(p_lq)^2\\
&+4(2b_0-5b_{l}-5b_{P}+3b_{ll}+6b_{lP}+3b_{PP})(p_Kp_l)m_l^2\\
&+2(-2b_0+5b_{l}+6b_{P}-b_{ll}-6b_{lP}-5b_{pp})(p_lq)m_l^2\\
&+8(-d_0+2d_{l}+2d_{P}-d_{ll}-2d_{lP}-d_{PP})(p_Kp_l)^2(p_lq)\\
&+2(-d_{l}-d_{P}+d_{ll}+2d_{lP}+d_{PP})(p_Kp_l)(p_lq)m_l^2\\
&+8(b_0-2b_{l}-3b_{P}+b_{ll}+3b_{lP}+2b_{PP})(p_Kp_l)(p_lq)\\
&+2(d_{P}+d_{ll}+d_{lP})(p_Kq)(p_lq)m_l^2-4b_2(m_l^2-2(p_Kp_l)+(p_lq))\\
&+4(-d_{K}+2d_{P}+d_{Kl}-2d_{lP}-2d_{PP})(p_Kp_l)(p_Kq)(p_lq));
\end{split}
\end{equation}

\begin{equation}
\begin{split}
|M|^2_{Low-3b-mag}&=\left(\frac{G}{\sqrt{2}}\sin{\theta_c}ef_+(0)\right)^2 \frac{e^2
\xi m^4_K}{\pi^2 (p_Kq)(p_lq)}\\ &\times
((b_{ll}+2b_{lP}+b_{PP})m_l^4-4b_2m_l^2+4(b_{PP}-b_{P})(p_lq)^2+8b_2(p_Kp_l)\\
&+4(b_{l}+b_{P}-2b_{ll}-4b_{lP}-2b_{PP})(p_Kp_l)m_l^2\\
&+4(b_{P}-b_{lP}-b_{PP})(p_Kq)m_l^2+2(-b_{l}-b_{P}+2b_{lP}+2b_{PP})(p_lq)m_l^2);
\end{split}
\end{equation}

\begin{equation}
\begin{split}
|M|^2_{Low-3b-Low}&=\left(\frac{G}{\sqrt{2}}\sin{\theta_c}ef_+(0)\right)^2
\frac{-2e^2 \xi m^4_K}{\pi^2(p_Kq)}\\ &\times
(d_{KK}(p_Kp_l)m_K^2-(d_{Kl}+d_{KP})m_l^2m_K^2+d_{Kl}(p_lq)m_K^2\\
&+(-2b_{l}-2b_{P}+b_{ll}+2b_{lP}+b_{PP})m_l^2\\
&+2(-d_{K}+d_{Kl}+d_{KP})(p_Kp_l)^2+2b_2+2d_2(p_Kp_l)\\
&+(-d_{ll}-2d_{lP}-d_{PP})(p_Kq)m_l^2\\
&+2d_{KP}(p_Kp_l)(p_Kq)+2(b_0-b_{l}-2b_{P}+b_{lP}+b_{PP})(p_lq)\\
&+2(-d_{l}-d_{P}+d_{ll}+2d_{lP}+d_{PP})(p_Kp_l)(p_lq).
\end{split}
\end{equation}


\begin{thebibliography}\\

\bibitem{gas}
J.\,Bijnens, G.\,Ecker, and J.\,Gasser,
\newblock Nucl.\,Phys.\,B 396, 81 (1993).

\bibitem{brag}
V.V.\,Braguta, A.A.\,Likhoded, A.E.\,Chalov,
\newblock Phys.\,Rev.\,D 65, 054038 (2002).

\bibitem{kub}
B.\,Kubis, E.H.\,Muller, J.\,Gasser, M.\,Schmid,
\newblock Eur.\,Phys.\,J. C 50, 557 (2007).

\bibitem{low}
F.E.\,Low,
\newblock Phys.\,Rev. 110, 974 (1958).

\bibitem{PDG08}
Particle Data Group,
\newblock Phys.\,Lett.\,B 667, 717 (2008).

\bibitem{PDG04}
Particle Data Group,
\newblock Phys.\,Lett.\,B 592, 619 (2004).

\bibitem{shim}
S.\,Shimizu {\em et al}.,
\newblock Phys.\,Lett.\,B 633, 190 (2006).

\bibitem{adler}
S.\,Adler {\em et al}.,
\newblock Phys.\,Rev.\,D 81, 092001 (2010).

\bibitem{akim}
S.A.\,Akimenko {\em et al}.,
\newblock Phys.\,Atom.\,Nucl. 70, 702 (2007).

\bibitem{mul}
E.H.\,Muller, B.\,Kubis,  U.-G.\,Meissner,
\newblock Eur.\,Phys.\,J. C 48, 427 (2006).

\bibitem{tchik}
O.G.\,Tchikilev {\em et al}.,
\newblock Phys.\,Atom.\,Nucl. 70, 29 (2007).

\end{thebibliography}
\end{document}